\documentclass[11pt,a4paper]{article}
\pdfoutput=1

\usepackage{amsmath,amssymb,tikz}
\usepackage{graphicx}
 \allowdisplaybreaks
\def\be{\begin{equation}}
\def\ee{\end{equation}}
\def\ba#1\ea{\begin{align}#1\end{align}}
\def\bg#1\eg{\begin{gather}#1\end{gather}}
\def\bm#1\em{\begin{multline}#1\end{multline}}
\def\bmd#1\emd{\begin{multlined}#1\end{multlined}}

\def\({\left(}
\def\){\right)}
\def\[{\left[}
\def\]{\right]}

\def \be {\begin{equation}}
\def \ee {\end{equation}}
\def \ba {\begin{array}}
\def \ea {\end{array}}
\def \bea{\begin{eqnarray}}
\def \eea{\end{eqnarray}}

\def\bea{\begin{eqnarray}}
\def\eea{\end{eqnarray}}

\newcommand{\bit}{\begin{itemize}}  \newcommand{\eit}{\end{itemize}}
\newcommand{\ben}{\begin{enumerate}}  \newcommand{\een}{\end{enumerate}}

\long\def\symbolfootnote[#1]#2{\begingroup%
\def\thefootnote{\fnsymbol{footnote}}\footnote[#1]{#2}\endgroup}

\newcommand{\sysu}{{\it School of Physics and Astronomy, Sun Yat-Sen University, 2 Daxue Road, Zhuhai 519082, China}}

\begin{document}
\thispagestyle{empty}
\begin{center}

~\vspace{20pt}

{\Large\bf Enhancement of Anomalous Boundary Current by High Temperature}

\vspace{25pt}

Ruiping Guo, Rong-Xin Miao ${}$\symbolfootnote[1]{Corresponding author email:~\sf
  miaorx@mail.sysu.edu.cn}

\vspace{10pt}${}$\sysu

\vspace{2cm}

\begin{abstract}
Recently it is found that Weyl anomaly leads to novel anomalous currents in the spacetime with a boundary.  However, the anomalous current is suppressed by the mass of charge carriers and the distance to the boundary, which makes it difficult to be measured. In this paper, we explore the possible mechanisms for the enhancement of anomalous currents. Interestingly, we find that the anomalous current can be significantly enhanced by the high temperature, which makes easier the experimental detection. For free theories, the anomalous current is proportional to the temperature in the high temperature limit. Note that the currents can be enhanced by thermal effects only at high temperatures. In general, this is not the case at low temperatures. For general temperatures, the absolute value of the current of  Neumann boundary condition first decreases and then increases with the temperature, while the current of Dirichlet boundary condition always increases with the temperature. It should be mentioned that the enhancement does not have an anomalous nature. In fact, the so-called anomalous current in this paper is not always related to Weyl anomaly. Instead, it is an anomalous effect due to the boundary.
\end{abstract}

\end{center}

\newpage
\setcounter{footnote}{0}
\setcounter{page}{1}

\tableofcontents

\section{Introduction}

Weyl anomaly measures the quantum violation of scaling symmetry of a theory, and has a wide range of applications in black-hole physics, cosmology and condensed matter physics \cite{Duff:1993wm}. It is interesting that Weyl anomaly is well-defined for not only conformal field theories but also the general quantum field theories \cite{Duff:1993wm,Brown:1976wc,Casarin:2018odz}. Recently, it is found that, due to Weyl anomaly, an external electromagnetic field can induce novel anomalous currents in a conformally flat space \cite{Chernodub:2016lbo, Chernodub:2017jcp} and a spacetime with boundaries \cite{Chu:2018ksb,Chu:2018ntx}.  It is similar to the anomaly-induced transport \cite{review} such as chiral magnetic effect (CME) \cite{Vilenkin:1995um,
Vilenkin:1980fu, Giovannini:1997eg, alekseev, Fukushima:2012vr} and chiral vortical effect (CVE) \cite{Kharzeev:2007tn,Erdmenger:2008rm,
 Banerjee:2008th,Son:2009tf,Landsteiner:2011cp,Golkar:2012kb,Jensen:2012kj}. See also \cite{Chu:2018fpx,Chu:2019rod,Miao:2017aba,Miao:2018dvm,Chernodub:2018ihb,Chernodub:2019blw,Ambrus:2019khr,Zheng:2019xeu,Miao:2018qkc,Chu:2020mwx,Chu:2020gwq,Hu:2020puq,Kawaguchi:2020kce,Kurkov:2020jet,Kurkov:2018pjw,Fialkovsky:2019rum,Cooper,Chu:2021eae} for related works. 

In this paper, we focus on the anomalous current in the spacetime with a boundary \cite{Chu:2018ksb,Chu:2018ntx}. It takes the following universal form in four dimensions
\begin{eqnarray}\label{typeIIcurrent}
 <{\bf J} > = \frac{e^2 c}{\hbar}
  \frac{4 b_1 {\bf n} \times {\bf B}}{x}+... , \quad x \ge \epsilon,
\end{eqnarray}
where $e$ is the electric charge, $c$ is the speed of light, $\hbar$ is the Planck constant, $b_1$ is a dimensionless constant, ${\bf n}$ is normal vector to the boundary, ${\bf B}$ is the external magnetic field,  $x$ is the proper distance to the boundary, $ \epsilon$ is a small cutoff of x and $...$ denotes higher order terms in $O(x)$.  It should be mentioned that there are boundary contributions to the current, which cancels the bulk ``divergence'' of 
(\ref{typeIIcurrent}) and makes finite the total current \cite{Chu:2018ksb}.  Note that (\ref{typeIIcurrent}) applies to the region near the boundary. The anomalous current in the full space is studied in \cite{Hu:2020puq}, where it is found that the mass and the distance to the boundary heavily suppress the currents. See also \cite{Chernodub:2018ihb}. 

In general dimensions, the anomalous currents take the following form  \cite{Chu:2018ksb,Chu:2018ntx}
\begin{eqnarray}\label{currentind}
 <J^d_{\mu}> \sim  \frac{F_{n \mu}}{x^{d-3}}+... , \quad x \ge \epsilon,
\end{eqnarray}
where $d$ denotes the dimension of spacetime, $F_{\mu\nu} $ is the field strength, $n$ is the normal direction to the boundary, and $...$ denotes sub-leading terms in $x$  .  In general, $J^d $ is not always related to Weyl anomaly. For example, there is no Weyl anomaly due to a background electromagnetic field $F_{\mu\nu}$ in two dimensions. However, following \cite{Chu:2018ksb,Chu:2018ntx}, we still call it ``anomalous current'' in this paper, since it can be regarded as an anomalous effect due to the boundary. Recall the fact that a constant magnetic field cannot produce non-zero renormalized currents in a flat space without boundaries.  Thus, the current (\ref{currentind}) is kind of ``anomalous'' in general dimensions.

In this paper,  we explore the mechanism for the enhancement of anomalous currents, which is important for the experimental measurement. We find that the high temperature can greatly enhance the anomalous currents. For free complex scalars, the anomalous current is proportional to the temperature in the high temperature limit. As a result, for any given mass and distance, one can always produce a detectable anomalous current by increasing the temperature.

Let us list our main results below. Recovering the units of anomalous current (\ref{4dJregularization}) in four dimensions, we have 
\begin{eqnarray}\label{currentIntrounit}
\lim_{T\to \infty}<{\bf J}> =\frac{k e^2 }{\hbar^2} T
\  {\bf n}  \times {\bf B}\  (c_0+ c_1 \ln (x \mu)), \quad x \sim 0,
\end{eqnarray}
where $k$ is Boltzmann constant, $T$ is the temperature, $\mu$ denotes the energy scale and $c_0, c_1$ are some dimensionless constants depending on the boundary conditions. Since the Planck constant appears in the denominator of (\ref{currentIntrounit}), it is clear that the anomalous current at high temperature is dominated by quantum effects.  In general dimensions, according to (\ref{novelcurrent}) we have 
\begin{eqnarray}\label{novelcurrentIntrounit}
\lim_{T\to \infty}<{\bf J}^d>= \frac{k }{\hbar c} T  \lim_{T\to 0} <{\bf J}^{d-1}>,
\end{eqnarray}
where $\lim_{T\to \infty}{\bf J}^d $ is the anomalous current at high temperature in $d$ dimensions, and $\lim_{T\to 0}{\bf J}^{d-1} $ is anomalous current at zero temperature in $(d-1)$ dimensions. It is remarkable that the anomalous currents at high temperature and zero temperature are related by (\ref{novelcurrentIntrounit}).  Since the anomalous current at zero temperature is a pure quantum effect, so does the the anomalous current at high temperatures. Interestingly,  the relation (\ref{novelcurrentIntrounit}) provides an indirect way to measure the anomalous current at zero temperature in lower dimensions, which can make easier the experiment.   Let us stress again that by ``anomalous current'', in general, we means the ``renormalized current" in a space with a boundary in this paper.  Besides, by ``enhancement of the current" we always means the absolute value of the current in this paper.  In general, the temperature can change the sign of the current.

Let us try to understand the interesting relation (\ref{novelcurrentIntrounit}). Usually one takes Euclidean signature to study the vacuum expectation value of currents at finite temperature.  The period of Euclidean time is given by $\beta=1/T$, which approaches zero in the high temperature limit.  According to Kaluza-Klein theory, a d-dimensional QFT with a small extra spacetime dimension (a small circle) behaves effectively as a (d-1)-dimensional QFT. That is why we could relate a d-dimensional current to a (d-1)-dimensional current in the high temperature limit.

Let us summarize the properties of anomalous currents at a finite temperature below. For simplicity, we focus on free complex scalars.
 
{\tt 1}.  The anomalous current is proportional to the temperature in the high temperature limit. Remarkably, the coefficient is just the anomalous current in lower dimensions at zero temperature. See (\ref{novelcurrent}). 

{\tt 2}.  The anomalous current of Dirichlet boundary condition (DBC) always increases with temperature, while the  absolute value of the current of Neumann boundary condition (NBC) first decreases and then increases with temperature.   See Fig. \ref{Figure5dcurrent} and Fig. \ref{Figure6dcurrent}.

{\tt 3}.  Although the large mass always suppresses anomalous currents, the small mass could enhance the anomalous current for NBC. See Fig. \ref{Figure6dmass} (right). 

The paper is organized as follows.
In section 2, by applying the heat-kernel method \cite{Vassilevich:2003xt,McAvity:1990we}, we study the anomalous current for free complex scalars at finite temperature up to the linear order of magnetic fields. We find that it is proportional to the temperature in the high temperature limit. In section 3, we obtain a non-perturbative formal expression of the anomalous current, which can be evaluated numerically. Again, the anomalous current is enhanced by the high temperature.  Finally, we conclude with some discussions in section 4.

\section{Anomalous current I: perturbative result}

In this section, by applying the heat kernel method \cite{Vassilevich:2003xt,McAvity:1990we}, we study the anomalous current of complex scalars at a finite temperature. It is found that, in the high temperature limit, the anomalous current increases linearly with temperature. As for the case of low temperature, the anomalous current increases with temperature for Dirichlet boundary condition (DBC), while decreases with temperature for Neumann boundary condition (NBC). 

For simplicity, we focus on a flat half space $x \ge 0$ with a constant magnetic field $B$ parallel to the boundary. We have coordinates $x^{\mu}=(\tau, x, y_a)=(\tau, x, y_1, ..., y_{d-2})$, background vector field $A_{\mu}=(0,0, B x, 0,..., 0)$ and the metric $g_{\mu\nu}=\delta_{\mu\nu}=\text{diag}(1,1,....,1)$. Here $\tau \simeq \tau+\beta$ is the Euclidean time,  $\beta=1/T$ is the inverse temperature and $x$ denotes the distance to the boundary. 

\subsection{Heat kernel}
The heat kernel of complex scalars satisfies the equation of motion (EOM)
 \begin{eqnarray}\label{EOM}
\partial_t K(t,x_{\mu},x'_{\mu})-\delta^{\mu\nu}(\partial_{\mu}+A_{\mu}) (\partial_{\nu}+A_{\nu})K(t,x_{\mu},x'_{\mu})=0
\end{eqnarray}
 together with the following boundary conditions (BC)
 \begin{eqnarray}\label{BCt}
&& \lim_{t\to 0} K(t,x_{\mu},x'_{\mu})=\delta^{d}(x_{\mu}-x'_{\mu}), \\
&& K(t,\tau, \tau')=K(t,\tau+\beta, \tau')=K(t,\tau, \tau'+\beta) \label{BCtau},
\end{eqnarray}
for $t$ and $\tau$, respectively. Besides, one further imposes either DBC
\begin{eqnarray}\label{DBC}
K(t,x_{\mu},x'_{\mu})|_{x=0}=0,
\end{eqnarray}
or NBC
\begin{eqnarray}
\partial_x K(t,x_{\mu},x'_{\mu})|_{x=0}=0, \label{NBC}
\end{eqnarray}
on the boundary  $x=0$. 

From the heat kernel, we can obtain the Green function
 \begin{eqnarray}\label{Greenfunction}
G(x_{\mu},x'_{\mu})=\int_0^{\infty} dt K(t,x_{\mu},x'_{\mu}),
\end{eqnarray}
and then derive the expectation value of the current by
\begin{eqnarray}\label{hatJ}
\hat{J}_{\mu}=\lim_{x'\to x} \left[ (\partial_{x_{\mu}}+A_{x_{\mu}})- (\partial_{x'_{\mu}}-A_{x'_{\mu}})\right] G(x_{\mu},x'_{\mu}).\nonumber\\
\end{eqnarray}
In general $\hat{J}_{\mu}$ is divergent, which can be renormalized by subtracting the value it would have in the space without boundary,
\begin{eqnarray}\label{renJ}
J_{\mu}=\hat{J}_{\mu}-\hat{J}_{0\mu}.
\end{eqnarray}

In general, it is difficult to solve the heat kernel in the spacetime with boundaries, even for free theories. For simplicity, we   focus on the perturbation solution in the linear order of the magnetic field $O(B)$. Following \cite{McAvity:1990we}, we obtain the heat kernel
\begin{eqnarray}\label{K}
K=\sum_{m=-\infty}^{\infty}\frac{1}{(4 \pi  t)^{\frac{d-3}{2}}\beta } \exp \left(-\frac{4 \pi ^2 m^2 t}{\beta ^2}+\frac{2 i \pi  m \left(\tau -\tau '\right)}{\beta }-\sum _{a=2}^{d-2} \frac{\left(y_a-y'_a\right){}^2}{4} \right)\big(K_0+K_{bdy}\big),
\end{eqnarray}
where $K_0$ is the heat kernel in a 2d free space
\begin{eqnarray}\label{K0}
K_0=\frac{B}{4 \pi  \sin (B t)} \exp \left(-\frac{B}{4}  \cot (B t)\left( (x-x')^2+(y_1-y'_1)^2\right)  +\frac{B}{2} \left(x'+x\right) \left(y'_1-y_1\right)\right),
\end{eqnarray}
and $K_{bdy}$ denotes the correction due to the boundary 
\begin{eqnarray}\label{Kbdy}
K_{bdy}=\frac{\chi B}{4 \pi  \sin (B t)} \exp \left(-\frac{B}{4}  \cot (B t)\left( (x+x')^2+(y_1-y'_1)^2\right) +\frac{B}{2} \left(x'+x +f_{BC}\right) \left(y'_1-y_1\right)\right).
\end{eqnarray}
Here $\chi=-1$ ($\chi=1$) for DBC (NBC) and $f_{BC}$ is given by
\begin{eqnarray}\label{fBC}
f_{BC}=\begin{cases}
\frac{-\sqrt{\pi } x x' e^{\frac{\left(x'+x\right)^2}{4 t}} \text{erfc}\left(\frac{x'+x}{2 \sqrt{t}}\right)}{\sqrt{t}}+O(B^2) , \ \ \ \ \ \ \ \ \ \ \ \ \ \ \ \ \ \ \ \ \ \ \ \ \  \ \ \ \ \ \ \ \  \text{DBC},\\
-(x'+x)+\frac{\sqrt{\pi } e^{\frac{\left(x'+x\right)^2}{4 t}} \left(2 t+x^2+\left(x'\right)^2\right) \text{erfc}\left(\frac{x'+x}{2 \sqrt{t}}\right)}{2 \sqrt{t}}+O\left(B^2\right), \ \ \ \ \ \ \text{NBC},
\end{cases}
\end{eqnarray}
where $ \text{erfc}(x)$ is the complementary error function. 
One can check that the heat kernel (\ref{K}) satisfies EOM (\ref{EOM}) and BCs (\ref{BCt},\ref{BCtau},\ref{DBC},\ref{NBC}) at the linear order of $O(B)$.

\subsection{Current at finite temperature}

Now we are ready to calculate the anomalous current at finite temperature. From (\ref{Greenfunction},\ref{hatJ},\ref{renJ},\ref{K}), we derive
\begin{eqnarray}\label{J1}
J_{y_1}=\sum_{m=-\infty}^{\infty} \int_0^{\infty} dt \frac{2 \pi  B}{  (4 \pi  t)^{d/2} \beta} e^{-\frac{4 \pi ^2 m^2 t}{\beta ^2}}\begin{cases}
 -x^2 \text{erfc}\left(\frac{x}{\sqrt{t}}\right) +O(B^2), \ \ \ \ \ \ \ \ \ \ \ \ \ \ \ \ \ \ \ \ \ \ \ \ \ \ \ \ \text{DBC},\\
\frac{2 \sqrt{t} x e^{-\frac{x^2}{t}}}{\sqrt{\pi }}-\left(t+x^2\right) \text{erfc}\left(\frac{x}{\sqrt{t}}\right)+O\left(B^2\right), \ \ \ \ \ \ \text{NBC}.
\end{cases}
\end{eqnarray}
Before we try to perform the above complicated sum and integral, let us first consider some interesting limits, the high temperature limit and low temperature limit.  In the high temperature limit $\beta\to 0$, only the term with $m=0$ contributes to the sum,
\begin{eqnarray}\label{highTsum}
\lim_{T\to \infty}\sum_{m=-\infty}^{\infty} e^{-\frac{4 \pi ^2 m^2 t}{\beta ^2}}= 1. 
\end{eqnarray}
Substituting (\ref{highTsum}) into (\ref{J1}) and performing the integral along t, we get
\begin{eqnarray}\label{highTJ1}
\lim_{T\to \infty}  J_{y_1}=\begin{cases}
-\frac{2^{2-d} \pi ^{\frac{1}{2}-\frac{d}{2}} \Gamma \left(\frac{d-1}{2}\right)}{(d-2) x^{d-4}} B \ T+O(B^2), \ \ \ \ \ \ \ \ \ \ \ \ \ \ \ \ \ \ \ \ \ \ \ \ \  \text{DBC},\\
\frac{2^{1-d} ((d-7) d+8) \pi ^{\frac{1}{2}-\frac{d}{2}} \Gamma \left(\frac{d-3}{2}\right)}{(d-4) (d-2) x^{d-4}} B\ T+O\left(B^2\right), \ \ \ \ \ \ \ \ \ \ \ \ \ \ \text{NBC}.
\end{cases}
\end{eqnarray}
It is remarkable that the anomalous current is proportional to the temperature in the high temperature limit. This provides an interesting mechanism to enhance the anomalous current, which is usually suppressed by the mass and the distance to the boundary.  Note that (\ref{highTJ1}) works for $d>2$ for DBC and $d>4$ for NBC.  For $d=4$, one should perform suitable regularization in order to get finite results. Taking the regularization \cite{Hu:2020puq}
\begin{eqnarray}\label{regularization}
J^{4d}_{y_1}=\lim_{\epsilon \to 0}\frac{ J_{y_1}(d=4+\epsilon)+J_{y_1}(d=4-\epsilon)}{2},
\end{eqnarray}
we derive the current in four dimensions
\begin{eqnarray}\label{4dJregularization}
\lim_{T\to \infty}J^{4d}_{y_1}=\begin{cases}
-\frac{1}{16 \pi }B\ T+O(B^2), \ \ \ \ \ \ \ \ \ \ \ \ \ \ \ \ \ \ \ \ \ \ \ \ \ \ \  \ \ \ \ \ \ \ \ \ \ \ \ \ \ \ \ \ \ \ \ \ \ \ \text{DBC},\\
\frac{4 \log (x)+3+\log (16)+2 \log (\pi )-2 \psi ^{(0)}\left(\frac{1}{2}\right)}{16 \pi } B\ T+O\left(B^2\right), \ \ \ \ \ \ \ \ \ \ \ \ \ \ \text{NBC},
\end{cases}
\end{eqnarray}
where $\psi^{(0)} $ is the PolyGamma function.

Let us go on to discuss the low temperature limit, where the discrete summation can be replaced by a continuous integration 
\begin{eqnarray}\label{lowTsum}
\lim_{T\to 0 }\sum_{m=-\infty}^{\infty} \frac{1}{\beta}e^{-\frac{4 \pi ^2 m^2 t}{\beta ^2}}= \int_{-\infty}^{\infty} e^{-4 \pi ^2 z^2 t}dz=\frac{1}{2 \sqrt{\pi  t}},
\end{eqnarray}
where $z=m/\beta$. Substituting (\ref{lowTsum}) into (\ref{J1}),  we derive
\begin{eqnarray}\label{lowTJ1}
\lim_{T\to 0}  J_{y_1}=\begin{cases}
-\frac{2^{1-d} \pi ^{-\frac{d}{2}} \Gamma \left(\frac{d}{2}\right)}{(d-1) x^{d-3}} B+O(B^2), \ \ \ \ \ \ \ \ \ \ \ \ \ \ \ \ \ \ \ \ \ \ \ \ \ \ \  \text{DBC},\\
\frac{2^{-d} \left(d^2-5 d+2\right) \pi ^{-\frac{d}{2}} \Gamma \left(\frac{d}{2}-1\right)}{(d-3) (d-1) x^{d-3}} B+O\left(B^2\right), \ \ \ \ \ \ \ \ \ \ \ \ \ \ \text{NBC},
\end{cases}
\end{eqnarray}
which agree with the results of \cite{Hu:2020puq,McAvity:1990we}.  This can be regarded as a check for our calculations. 

There is another method to study the current in the high and low temperature limits. By applying the transformation 
\begin{eqnarray}\label{transformation}
\sum_{m=-\infty}^{\infty}\frac{e^{-\frac{4 \pi ^2 m^2 t}{\beta ^2}}}{\beta }=\sum_{m=-\infty}^{\infty}\frac{e^{-\frac{\beta ^2 m^2}{4 t}}}{2 \sqrt{\pi } \sqrt{t}},
\end{eqnarray}
we can rewrite the current (\ref{J1}) into the following form
\begin{eqnarray}\label{J2}
J_{y_1}=\sum_{m=-\infty}^{\infty} \int_0^{\infty} dt \frac{2 \pi  B}{  (4 \pi  t)^{(d+1)/2} } e^{-\frac{\beta ^2 m^2 }{4t}}\begin{cases}
 -x^2 \text{erfc}\left(\frac{x}{\sqrt{t}}\right) +O(B^2), \ \ \ \ \ \ \ \ \ \ \ \ \ \ \ \ \ \ \ \ \ \ \ \ \ \ \ \ \text{DBC},\\
\frac{2 \sqrt{t} x e^{-\frac{x^2}{t}}}{\sqrt{\pi }}-\left(t+x^2\right) \text{erfc}\left(\frac{x}{\sqrt{t}}\right)+O\left(B^2\right), \ \ \ \ \ \ \text{NBC}.
\end{cases}
\end{eqnarray}
Note that the transformation (\ref{transformation}) maps the high (low) temperature to the low (high) temperature. 
Now the term with $m=0$ dominates the sum in the low temperature limit
\begin{eqnarray}\label{lowTsum1}
\lim_{T\to 0 }\sum_{m=-\infty}^{\infty}\frac{e^{-\frac{\beta ^2 m^2}{4 t}}}{2 \sqrt{\pi  t} }=\frac{1}{2 \sqrt{\pi  t}},
\end{eqnarray}
which exactly agrees with (\ref{lowTsum},\ref{transformation}).  Substituting (\ref{lowTsum1}) into (\ref{J2}),  we re-derive (\ref{lowTJ1}).  As in the high temperature limit, the sum can be replaced by the following integral
\begin{eqnarray}\label{highTsum1}
\lim_{T\to \infty }\sum_{m=-\infty}^{\infty} \frac{e^{-\frac{\beta ^2 m^2}{4 t}}}{2 \sqrt{\pi } \sqrt{t}}= \int_{-\infty}^{\infty} \frac{e^{-\frac{\bar{z}^2}{4 t}}}{2 \sqrt{\pi } \sqrt{t}\beta}d\bar{z}=\frac{1}{\beta},
\end{eqnarray}
which agrees with (\ref{highTsum},\ref{transformation}). From (\ref{J2},\ref{highTsum1}), we reproduce the anomalous current (\ref{highTJ1}) in high temperature limit.  Now we have obtained the currents in the low and high temperature limits by using two methods. This is a double check for our calculations. 

\begin{figure}[t]
\centering
\includegraphics[width=7.5cm]{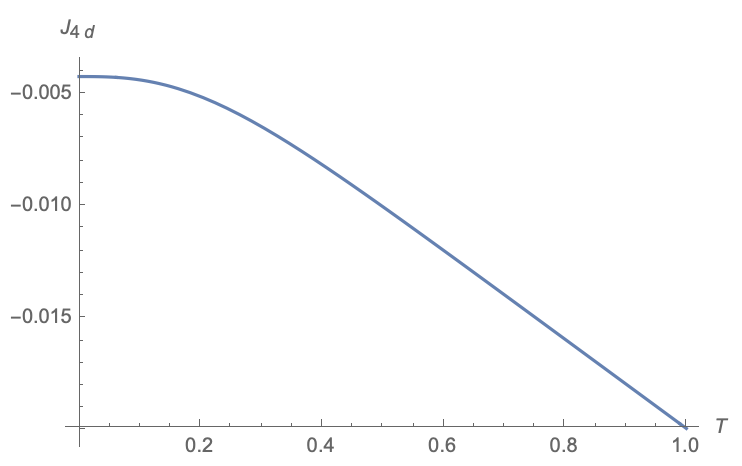}
\includegraphics[width=7.5cm]{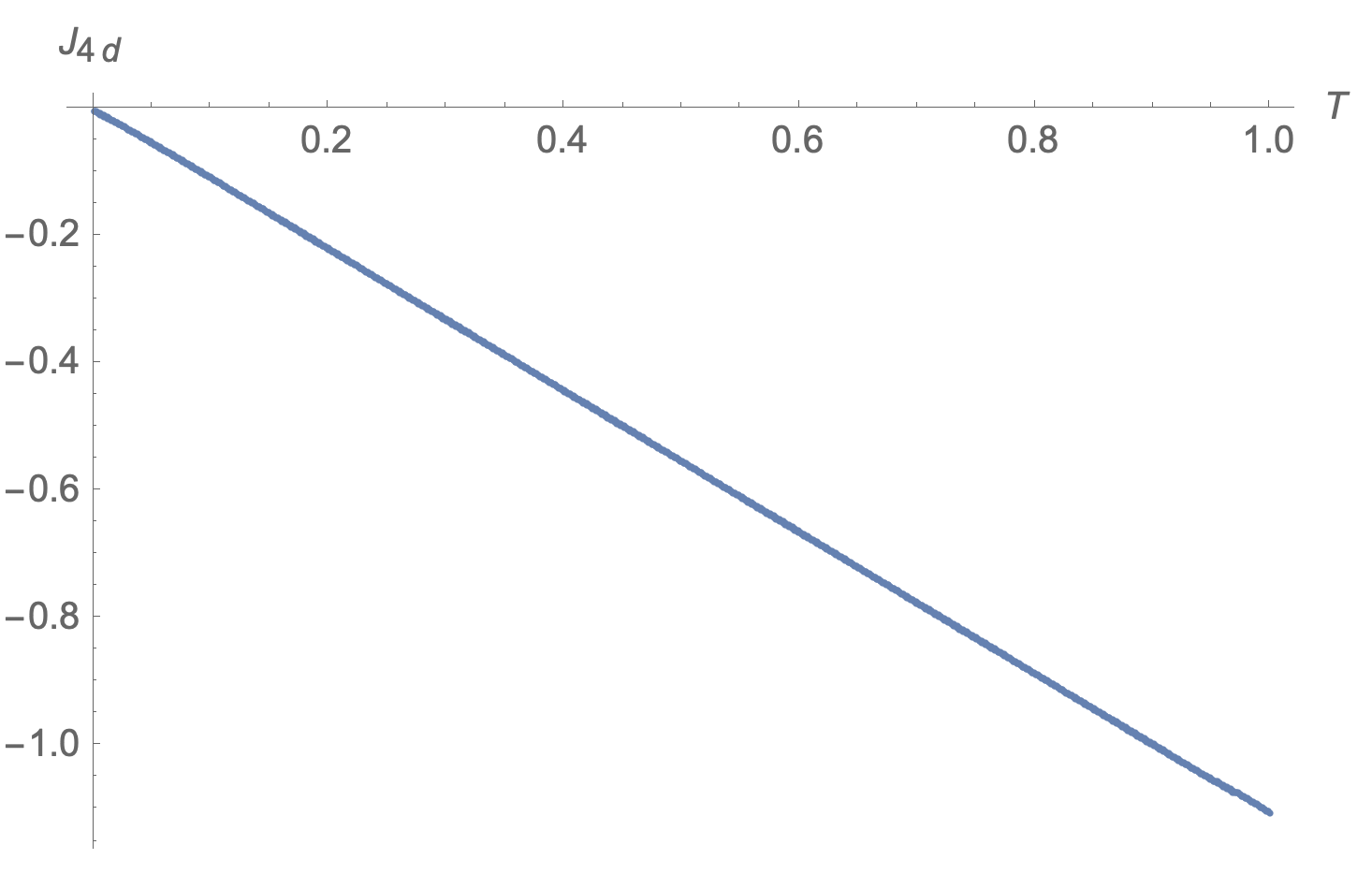}
\caption{Anomalous currents of complex scalars for DBC (left) and NBC (right) in four dimensions. The current of NBC is much larger than the current of DBC. Here we have set $B=x=1$.}
\label{Figure4dcurrent}
\end{figure}

\begin{figure}[]
\centering
\includegraphics[width=10cm]{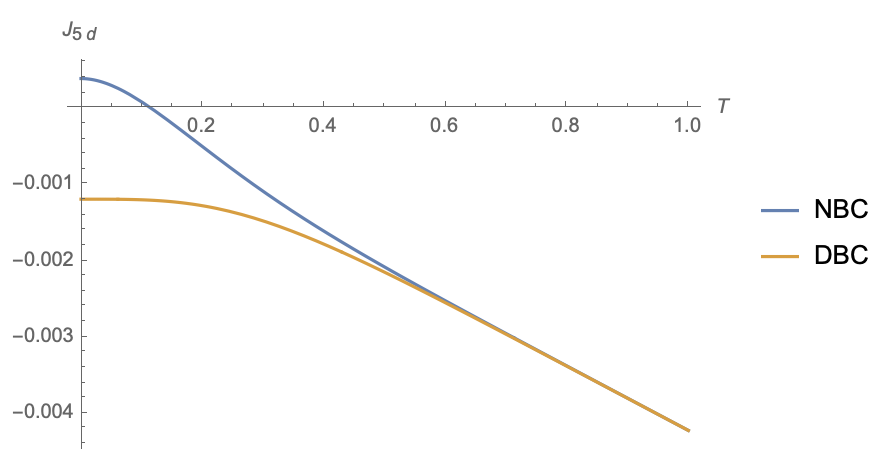}
\caption{Anomalous currents of complex scalars for DBC (orange line) and NBC (blue line) in five dimensions. It is remarkable that the currents increase linearly with temperature and approach the same value for DBC and NBC in the high temperature limit.  }
\label{Figure5dcurrent}
\end{figure}

\begin{figure}[t]
\centering
\includegraphics[width=7.5cm]{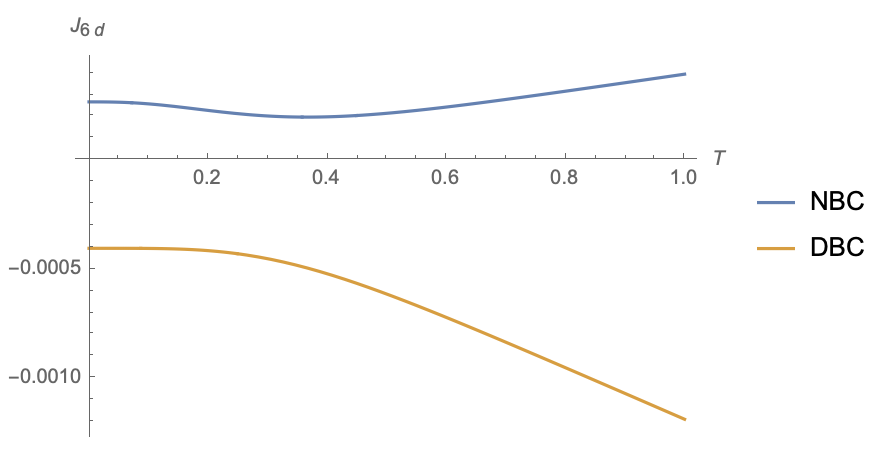}
\includegraphics[width=7.5cm]{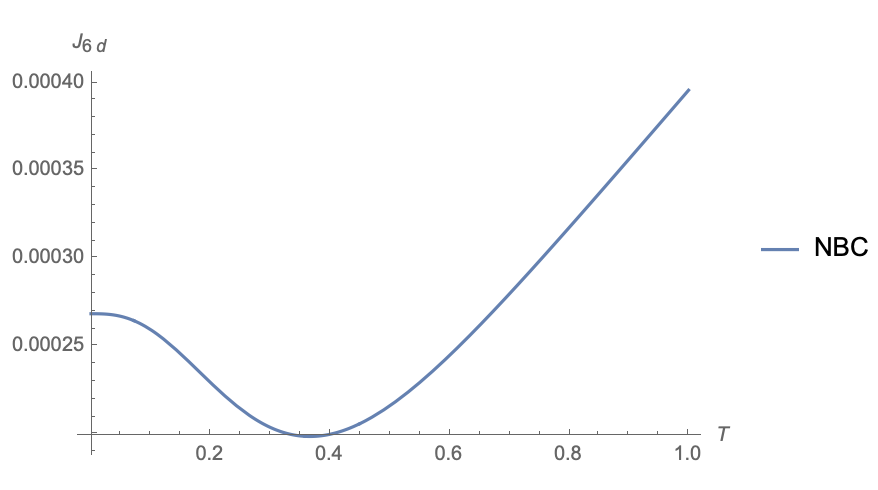}
\caption{Anomalous currents of complex scalars for DBC (orange line) and NBC (blue line) in six dimensions. The current of NBC first decreases and then increases with temperature, while the current of DBC always increases with temperature. }
\label{Figure6dcurrent}
\end{figure}

Let us go on to consider the general temperature. Summing (\ref{J2}), we get
\begin{eqnarray}\label{J3}
J_{y_1}= \int_0^{\infty} dt \frac{2 \pi  B \vartheta _3\left(0,e^{-\frac{\beta ^2}{4 t}}\right)}{  (4 \pi  t)^{(d+1)/2} } \begin{cases}
 -x^2 \text{erfc}\left(\frac{x}{\sqrt{t}}\right) +O(B^2), \ \ \ \ \ \ \ \ \ \ \ \ \ \ \ \ \ \ \ \ \ \ \ \ \ \ \ \ \text{DBC},\\
\frac{2 \sqrt{t} x e^{-\frac{x^2}{t}}}{\sqrt{\pi }}-\left(t+x^2\right) \text{erfc}\left(\frac{x}{\sqrt{t}}\right)+O\left(B^2\right), \ \ \ \ \ \ \text{NBC},
\end{cases}
\end{eqnarray}
where $ \vartheta _3$ is the Elliptic theta function.  Although it is difficult to work out the exact expression of (\ref{J3}), it can be evaluated numerically.  See Fig.\ref{Figure4dcurrent}, Fig.\ref{Figure5dcurrent}  and Fig.\ref{Figure6dcurrent}  for examples. Without loss of generality, we set $B=x=1$ for all the figures of this paper. We find that, in the high temperature limit, the currents increase linearly with temperature for both DBC and NBC in general dimensions. It is interesting that the currents of DBC and NBC approach the same high-temperature limit in five dimensions. It is also interesting that, in dimensions higher than four,  the absolute values of the currents of NBC first decrease and then increase with temperature, while the currents of DBC always increase with temperature. 

\subsection{Mass effect}

We focus on massless scalars in the above discussions, where exact expressions of the currents can be derived in the high and low temperature limits.  In this subsection, let us study the mass effect, which can be taken into account by adding $e^{-M^2 t}$ to the heat kernel (\ref{K}). Following the approach of sect. 2.2, we obtain the anomalous currents for massive scalars as
\begin{eqnarray}\label{J4}
J_{y_1}= \int_0^{\infty} dt \frac{2 \pi  B \vartheta _3\left(0,e^{-\frac{\beta ^2}{4 t}}\right)e^{-M^2 t}}{  (4 \pi  t)^{(d+1)/2} }  \begin{cases}
 -x^2 \text{erfc}\left(\frac{x}{\sqrt{t}}\right) +O(B^2), \ \ \ \ \ \ \ \ \ \ \ \ \ \ \ \ \ \ \ \ \ \ \ \ \ \ \ \ \text{DBC},\\
\frac{2 \sqrt{t} x e^{-\frac{x^2}{t}}}{\sqrt{\pi }}-\left(t+x^2\right) \text{erfc}\left(\frac{x}{\sqrt{t}}\right)+O\left(B^2\right), \ \ \ \ \ \ \text{NBC},
\end{cases}
\end{eqnarray}
where $M$ is the scalar mass.  It is clear that the current (\ref{J4}) is heavily suppressed by large mass.  
In the high and low temperature limits, the anomalous current (\ref{J4}) becomes
\begin{eqnarray}\label{J4highT}
\lim_{T\to \infty}J_{y_1}= T\int_0^{\infty} dt \frac{2 \pi  B}{  (4 \pi  t)^{d/2} } e^{-M^2 t}\begin{cases}
 -x^2 \text{erfc}\left(\frac{x}{\sqrt{t}}\right) +O(B^2), \ \ \ \ \ \ \ \ \ \ \ \ \ \ \ \ \ \ \ \ \ \ \ \ \ \ \ \ \text{DBC},\\
\frac{2 \sqrt{t} x e^{-\frac{x^2}{t}}}{\sqrt{\pi }}-\left(t+x^2\right) \text{erfc}\left(\frac{x}{\sqrt{t}}\right)+O\left(B^2\right), \ \ \ \ \ \ \text{NBC},
\end{cases}
\end{eqnarray}
and 
\begin{eqnarray}\label{J4highT}
\lim_{T\to 0 }J_{y_1}= \int_0^{\infty} dt \frac{2 \pi  B}{  (4 \pi  t)^{(d+1)/2} } e^{-M^2 t}\begin{cases}
 -x^2 \text{erfc}\left(\frac{x}{\sqrt{t}}\right) +O(B^2), \ \ \ \ \ \ \ \ \ \ \ \ \ \ \ \ \ \ \ \ \ \ \ \ \ \ \ \ \text{DBC},\\
\frac{2 \sqrt{t} x e^{-\frac{x^2}{t}}}{\sqrt{\pi }}-\left(t+x^2\right) \text{erfc}\left(\frac{x}{\sqrt{t}}\right)+O\left(B^2\right), \ \ \ \ \ \ \text{NBC}.
\end{cases}
\end{eqnarray}
It is remarkable that, similar to the massless case, the anomalous currents of massive scalars also increase linearly with temperature in the high temperature limit. This means that, for a given charge carrier with fixed mass, by increasing the temperature, one can always produce a detectable anomalous current in laboratory. 

\begin{figure}[t]
\centering
\includegraphics[width=7.5cm]{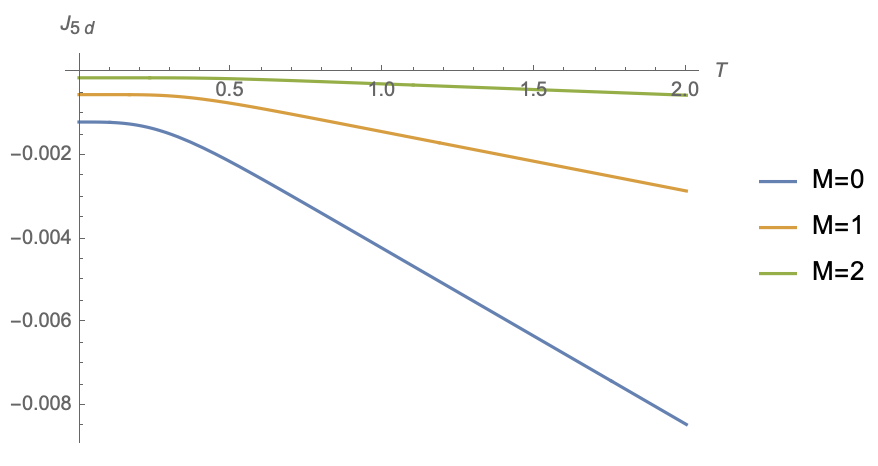}
\includegraphics[width=7.5cm]{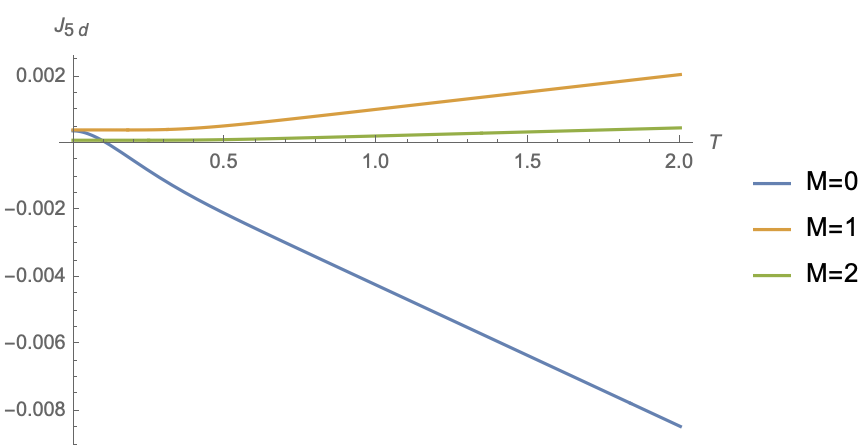}
\caption{Mass effects of currents for DBC (left) and NBC (right) in five dimensions. It is interesting that the mass can change the sign of currents of NBC.  }
\label{Figure5dmass}
\end{figure}

\begin{figure}[t]
\centering
\includegraphics[width=7.5cm]{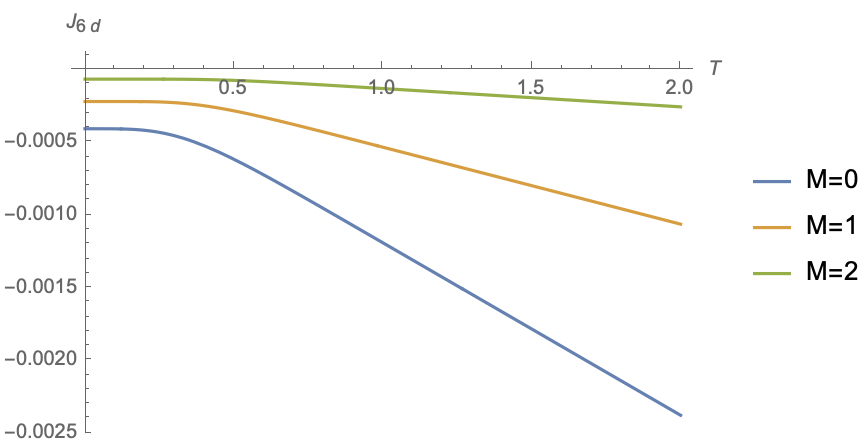}
\includegraphics[width=7.5cm]{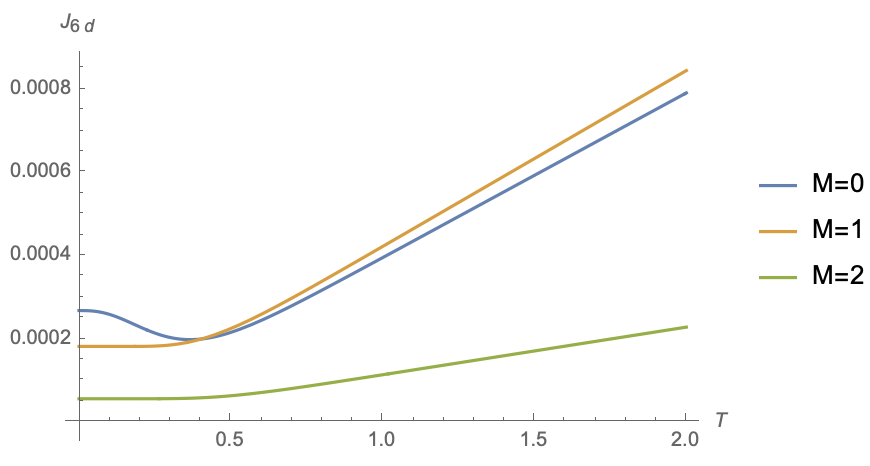}
\caption{Mass effects of currents for DBC (left) and NBC (right) in six dimensions. It is remarkable that a suitable small mass can enhance the currents of NBC.  }
\label{Figure6dmass}
\end{figure}

\begin{figure}[t]
\centering
\includegraphics[width=7.5cm]{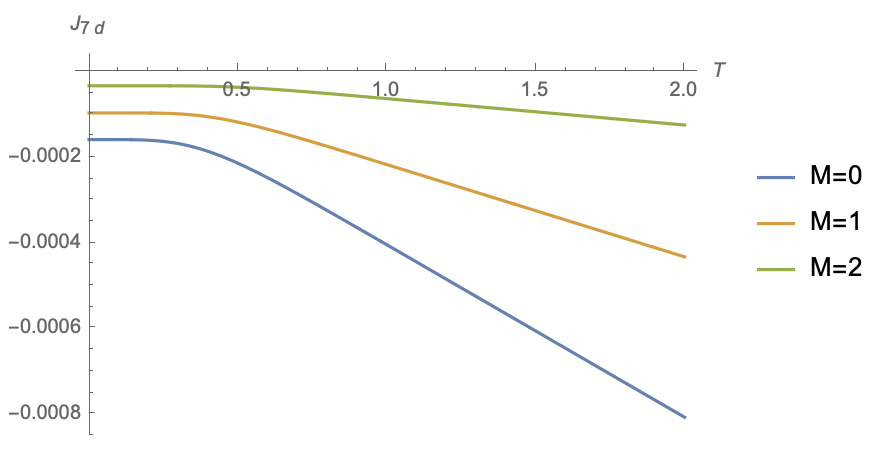}
\includegraphics[width=7.5cm]{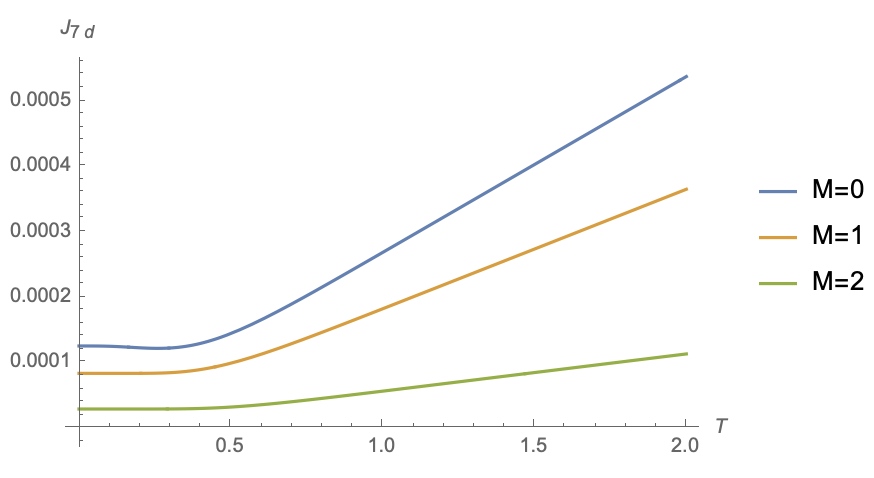}
\caption{Mass effects of currents for DBC (left) and NBC (right) in seven dimensions. The mass suppresses the currents of DBC and NBC.  }
\label{Figure7dmass}
\end{figure}

To end this section, let us draw some figures to illustrate the mass effect of anomalous currents. See Figs. \ref{Figure5dmass},\ref{Figure6dmass},\ref{Figure7dmass}. In general, the large mass suppresses but does not change the high-temperature behaviors of the currents. In other words, the currents are always enhanced by high temperatures. It is remarkable that, as is shown in figure 4,  the current of NBC can also be enhanced by increasing the mass slightly.

\section{Anomalous current II: exact result}

In the above section, we focus on the currents at the linear order of the magnetic field $O(B)$. In this section, we generalize our discussions to non-perturbative currents.  We follow the method of \cite{Hu:2020puq}, where the anomalous current at zero temperature is investigated. 

Green's function obeys EOM
\begin{eqnarray}\label{GreenEOMexact}
[-D^{\mu}D_{\mu}+M^2 ]G(x,x')=\delta^{(d)}(x,x'),
\end{eqnarray}
and DBC
\begin{eqnarray}\label{GreenDBC}
G(x_{\mu},x'_{\mu})|_{x=0}=0,
\end{eqnarray}
or NBC
\begin{eqnarray}
\partial_x G(x_{\mu},x'_{\mu})|_{x=0}=0, \label{GreenNBC}
\end{eqnarray}
on the boundary  $x=0$. Performing Fourier transform with the correct period $\tau\simeq \tau+\beta$, 
\begin{eqnarray}\label{Fouriertransform}
G=\frac{1}{ \beta}\sum_{m=-\infty}^{\infty}\int \frac{dk_{\parallel}^{d-2}}{(2\pi)^{d-2}}\tilde{G}(k,m) e^{i \frac{2\pi m }{\beta}(\tau-\tau')}e^{-i k_{a}(y_{a}-y'_a)}
\end{eqnarray}
we can rewrite (\ref{GreenEOMexact}) as
\begin{eqnarray}\label{GreenEOMexact1}
[-\partial_x^2+(M^2+k^2_a+(\frac{2\pi m}{\beta})^2) -2 B x k_{1}+ B^2 x^2]\tilde{G}=\delta(x-x'),
\end{eqnarray}
where $k_1=k_{y_1}$. We split Green's function $\tilde{G}$ into the one in a free space and the correction due to the boundary
\begin{eqnarray}\label{Gk}
\tilde{G}=G_{free}+ G_{bdy},
\end{eqnarray}
where $G_{free}$ is given by \cite{JohnP,FalKovskii}
\begin{eqnarray}\label{Gfree}
G_{free}=\begin{cases}\sqrt{\frac{1}{4 \pi B}}  \Gamma (\lambda_k ) D_{-\lambda_k }\left(\sqrt{2} \left(\bar{x}-\bar{k}_1\right)\right) D_{-\lambda_k }\left(\sqrt{2} \left(\bar{k}_1-\bar{x'}\right)\right)
,
\ \ \ \ \ \ \ \ x> x',\\
\sqrt{\frac{1}{4 \pi B}}  \Gamma (\lambda_k ) D_{-\lambda_k }\left(\sqrt{2} \left(\bar{k}_1-\bar{x}\right)\right) D_{-\lambda_k }\left(\sqrt{2} \left(\bar{x'}-\bar{k}_1\right)\right)
,
\ \ \ \ \ \ \ \ x<x'.
\end{cases}
\end{eqnarray}
Here $D$ denotes the parabolic cylinder function, $\lambda_k=(B+M^2+k^2_a+(\frac{2\pi m}{\beta})^2-k_1^2)/(2 B)$, $\bar{k}_1=k_1/\sqrt{B}$ and $\bar{x}=\sqrt{B}x$.   Imposing BCs (\ref{GreenDBC}, \ref{GreenNBC}), we solve the corrections to Green's function
\begin{eqnarray}\label{GbdyDBC}
G_{bdy}=\frac{-\Gamma \left(\lambda _k\right) D_{-\lambda _k}\left(\sqrt{2} \bar{k}_1\right) }{2 \pi ^{1/2} \sqrt{B} D_{-\lambda _k}\left(-\sqrt{2} \bar{k}_1\right)} D_{-\lambda _k}\left(\sqrt{2} \left( \bar{x}-\bar{k}_1\right)\right)D_{-\lambda _k}\left(\sqrt{2} \left(\bar{x'}-\bar{k}_1\right)\right)
\end{eqnarray}
for DBC and 
\begin{eqnarray}\label{GbdyNBC}
G_{bdy}=\frac{\Gamma \left(\lambda _k\right) \left(\sqrt{2} D_{1-\lambda _k}\left(\sqrt{2} \bar{k}_1\right)- \bar{k}_1 D_{-\lambda _k}\left(\sqrt{2} \bar{k}_1\right)\right)}{2 \pi ^{1/2} \sqrt{B} \left(\sqrt{2} D_{1-\lambda _k}\left(-\sqrt{2} \bar{k}_1\right)+\bar{k}_1 D_{-\lambda _k}\left(-\sqrt{2} \bar{k}_1\right)\right)} D_{-\lambda _k}\left(\sqrt{2} \left( \bar{x}-\bar{k}_1\right)\right)D_{-\lambda _k}\left(\sqrt{2} \left(\bar{x'}-\bar{k}_1\right)\right)\nonumber\\
\end{eqnarray}
for NBC. 

Now we are ready to derive the anomalous current. Substituting (\ref{Fouriertransform},\ref{GbdyDBC},\ref{GbdyNBC}) into (\ref{hatJ},\ref{renJ}) , we get the renormalized current
\begin{eqnarray}\label{CurrentDBC}
J_{y_1}=\frac{-1}{ \beta}\sum_{m=-\infty}^{\infty}\int_{-\infty}^{\infty} dp^{d-3}dk_1\frac{\left(\bar{x}-\bar{k}_1\right) \Gamma \left(\lambda _p\right) D_{-\lambda _p}\left(\sqrt{2} \bar{k}_1\right)}{2^{d-2} \pi ^{d-\frac{3}{2}} D_{-\lambda _p}\left(-\sqrt{2} \bar{k}_1\right)}D_{-\lambda _p}\left(\sqrt{2} \left(\bar{x}-\bar{k}_1\right)\right){}^2,
\end{eqnarray}
for DBC and 
\begin{eqnarray}\label{CurrentNBC}
&&J_{y_1}=\frac{1}{\beta}\sum_{m=-\infty}^{\infty}\int_{-\infty}^{\infty}dp^{d-3}dk_1 \frac{(\bar{x}-\bar{k}_1) \Gamma \left(\lambda _p\right) \left(\sqrt{2} D_{1-\lambda _p}\left(\sqrt{2} \bar{k}_1\right)-\bar{k}_1 D_{-\lambda _p}\left(\sqrt{2} \bar{k}_1\right)\right)}{2^{d-2} \pi ^{d-\frac{3}{2}} \left(\sqrt{2} D_{1-\lambda _p}\left(-\sqrt{2} \bar{k}_1\right)+\bar{k}_1 D_{-\lambda _p}\left(-\sqrt{2} \bar{k}_1\right)\right)}\nonumber\\
&&\ \ \ \ \ \ \ \ \ \ \ \ \ \times D_{-\lambda _p}\left(\sqrt{2} \left(\bar{x}-\bar{k}_1\right)\right){}^2,
\end{eqnarray}
for NBC. Recall that $\lambda_p=(B+M^2+(\frac{2\pi m}{\beta})^2+p^2)/(2 B)$, $\bar{k}_1=k_1/\sqrt{B}$ and $\bar{x}=\sqrt{B}x$.  In principle, the formal expressions (\ref{CurrentDBC},\ref{CurrentNBC}) can be evaluated numerically.  

In the low temperature limit $\beta \to \infty$, the sum can be replaced by the integral 
\begin{eqnarray}\label{sumintagralexact}
\frac{1}{ \beta}\sum_{m=-\infty}^{\infty} =\int_{-\infty}^{\infty} \frac{dp_{\tau}}{2\pi},
\end{eqnarray}
where $p_{\tau}=2\pi m/\beta$. And the currents (\ref{CurrentDBC},\ref{CurrentNBC}) reduce to exactly the ones at zero temperature \cite{Hu:2020puq}
\begin{eqnarray}\label{CurrentDBClowT}
\lim_{T\to 0}J_{y_1}=-\int_{-\infty}^{\infty} dp^{d-2}dk_1\frac{\left(\bar{x}-\bar{k}_1\right) \Gamma \left(\lambda _p\right) D_{-\lambda _p}\left(\sqrt{2} \bar{k}_1\right)}{2^{d-1} \pi ^{d-\frac{1}{2}} D_{-\lambda _p}\left(-\sqrt{2} \bar{k}_1\right)}D_{-\lambda _p}\left(\sqrt{2} \left(\bar{x}-\bar{k}_1\right)\right){}^2,
\end{eqnarray}
for DBC and 
\begin{eqnarray}\label{CurrentNBClowT}
&&\lim_{T\to 0}J_{y_1}=\int_{-\infty}^{\infty}dp^{d-2}dk_1 \frac{(\bar{x}-\bar{k}_1) \Gamma \left(\lambda _p\right) \left(\sqrt{2} D_{1-\lambda _p}\left(\sqrt{2} \bar{k}_1\right)-\bar{k}_1 D_{-\lambda _p}\left(\sqrt{2} \bar{k}_1\right)\right)}{2^{d-1} \pi ^{d-\frac{1}{2}} \left(\sqrt{2} D_{1-\lambda _p}\left(-\sqrt{2} \bar{k}_1\right)+\bar{k}_1 D_{-\lambda _p}\left(-\sqrt{2} \bar{k}_1\right)\right)}\nonumber\\
&&\ \ \ \ \ \ \ \ \ \ \ \ \ \times D_{-\lambda _p}\left(\sqrt{2} \left(\bar{x}-\bar{k}_1\right)\right){}^2,
\end{eqnarray}
for NBC, where $\lambda_p=(B+M^2+p_a^2+p_{\tau}^2)/(2 B)$.

Note that only the combination $M^2+(\frac{2\pi m}{\beta})^2$ appears in the currents (\ref{CurrentDBC},\ref{CurrentNBC}). Thus $(\frac{2\pi m}{\beta})$ behaves effectively as a mass. In the high temperature limit, the effective mass  $M_{eff}^2=M^2+(\frac{2\pi m}{\beta})^2$ becomes infinite for $m\ne 0$, which would heavily suppress the current \footnote{One can check that the integrand functions of (\ref{CurrentDBC}) and (\ref{CurrentNBC}) approach zero as $M_{eff}\to \infty$.}. As a result, only the term with $m=0$ is dominated in the high temperature limit.  Keeping only the zero-m terms of (\ref{CurrentDBC},\ref{CurrentNBC}) and using (\ref{CurrentDBClowT},\ref{CurrentNBClowT}), we finally obtain the anomalous current in the high temperature limit
\begin{eqnarray}\label{novelcurrent}
&&\lim_{T\to \infty}J^d_{y_1}= T  \lim_{T\to 0} J^{d-1}_{y_1},
\end{eqnarray}
where $J^d$ denotes the current in d dimensions.  It is remarkable that the non-perturbative current is also proportional to the temperature in the high temperature limit. It is also remarkable that the coefficient is just the current in ($d-1$) dimensions at zero temperature. One can check that the perturbative currents obtained in sect. 2 obey the novel relation (\ref{novelcurrent}).

\section{Conclusions and Discussions}

In this paper, we explore the mechanism to enhance the anomalous current caused by a background magnetic field in the spacetime with a boundary.  Usually, the anomalous current is suppressed by the mass and the distance to the boundary, which are the main experimental obstructions.  Remarkably, we find that the high temperature can greatly enhance the anomalous current and make easier the experimental measurement.  For free complex scalars, it is found that  the anomalous current is proportional to the temperature in the high temperature limit. Interestingly, the coefficient is just the current in lower dimensions at zero temperature. Thus, for any given charge carrier with a fixed mass $M$, one can always produce a detectable anomalous current by increasing the temperature. We look forward to the experimental detection of this novel anomalous current. For simplicity, we focus on free complex scalars in this paper. It is interesting to generalize the results of this paper to Dirac fields. It is also interesting to study the holographic anomalous current at finite temperature following the approach of \cite{Chu:2018ntx,Miao:2018qkc,Liu:2021lbh}.  Note that (\ref{novelcurrentIntrounit}) shows that the anomalous current at zero temperature in four dimensions is related to the renormalized current at high temperature in five dimension. This implies that there is an ``effective Weyl anomaly" in the high temperature limit in five dimensions, which is consistent with the Kaluza-Klein mechanism. According to the Kaluza-Klein theory, a 5-dimensional Euclidean QFT with a small period of Euclidean time $\beta=1/T$ behaves effectively as a 4-dimensional Euclidean QFT, which is expected to has a Weyl anomaly. We hope these problems could be addressed in future.

\section*{Acknowledgements}
R. X. Miao acknowledges the supports from Guangdong Basic and Applied Basic Research Foundation (No.2020A1515010900) and NSFC grant (No. 11905297). 

\appendix

\end{document}